\documentclass[preprint,showpacs,preprintnumbers,
amsmath,amssymb]{revtex4}

\usepackage{graphicx}% Include figure files
\usepackage{dcolumn}% Align table columns on decimal point
\usepackage{bm,morefloats,color}% bold math

\usepackage{amssymb,bm}

\def\Eq#1{\begin{equation} #1 \end{equation}}
\def\Eqr#1{\begin{eqnarray} #1 \end{eqnarray}}
\def\Eqrsubl#1#2{\begin{subequations}\label{#1}\Eqr{#2}\end{subequations}}

\newcommand{\nn}{\nonumber}
\newcommand{\pd}{\partial}

\newcommand{\bea}{\begin{eqnarray}}
\newcommand{\eea}{\end{eqnarray}}

\def\Xsp{{\rm X}}

\def\Ysp{{\rm Y}}
\def\Zsp{{\rm Z}}
\def\X5sp{{\rm X}_5}
\def\Y3sp{{\rm Y}_3}
\def\Z3sp{{\rm Z}_3}

\def\na{{\nabla}}

\def\defscript{\mathscr}

\def\X{{\defscript X}}
\def\Y{{\defscript Y}}
\def\Z{{\defscript Z}}

\def\linebreak{\hfill\break}

\def\bra<#1|{\langle #1\rvert}
\def\ket|#1>{\lvert#1 \rangle}
\def\braket<#1|#2>{\langle #1|#2 \rangle}

\def\const{\text{const}}
\def\otop#1{\hbox{$#1\kern-0.1em$\llap{\hbox{\raise1.7ex\hbox{$\scriptstyle\circ$}}}} }

\def\bigpare(#1){\left(#1\right)}

\def\bigbra[#1]{\left[ #1 \right]}

\def\therefore{\mbox{\setbox0=\hbox{X}\hbox{$\ldotp$}\raise0.7\ht0\hbox{$\ldotp$}\hbox{$\ldotp$}} \quad }
\def\because{\mbox{\setbox0=\hbox{X}\raise0.7\ht0\hbox{$\ldotp$}\hbox{$\ldotp$}\raise0.7\ht0\hbox{$\ldotp$}}\kern0pt }

\def\Frac(#1/#2){\left(\frac{#1}{#2}\right)}

\def\pd{\partial}
\def\AdS{{\rm AdS}}

\def\Eq#1{\begin{equation} #1 \end{equation}}

\def\Eqr#1{\begin{eqnarray} #1 \end{eqnarray}}

\def\Eqrsubl#1#2{\begin{subequations}\label{#1}\Eqr{#2}\end{subequations}}
\def\Bitm{\begin{itemize}}
\def\Eitm{\end{itemize}}
\def\Blist#1#2{\begin{list}{#1}{\parsep=0pt \itemsep=0pt%   
  \listparindent=0pt #2}}
\def\Elist{\end{list}}
\long\def\ignore#1#2{\def\ignoreflag{#1}\long\def\tmptext{#2}
  \ifnum\ignoreflag>1 #2 \fi}

\begin{document}

\title{Supersymmetry in a dynamical M-brane background}% Force line breaks with \\

\author{Kengo Maeda}%
\affiliation{%
Faculty of Engineering, 
Shibaura Institute of Technology, 
Saitama 330-8570, Japan}%

\author{Kunihito Uzawa}
\affiliation{%
Department of Physics,
School of Science and Technology,
Kwansei Gakuin University, Sanda, Hyogo 669-1337, Japan
}%

\date{\today}% It is always \today, today,

\begin{abstract}
The supersymmetry arises in certain theories of fermions coupled to 
gauge fields and gravity in a spacetime of 11 dimensions. 
 The dynamical brane background has mainly been studied for the class of
purely bosonic solutions only, but recent developments involving 
a time-dependent brane solution have made it clear that 
one can get more information by asking what happens on 
supersymmetric systems.
In this paper, we construct an exact supersymmetric solution of a 
dynamical M-brane background in the 11-dimensional supergravity and 
investigate supersymmetry breaking, the geometric features near 
the singularity and the black hole horizon. 
\end{abstract}

\pacs{04.65.+e, 11.27.+d, 11.25.-w, 04.50.Gh, 04.20.Dw}

\maketitle

%======================================%
%<<<<<<<<<<<<< SECTION 1 >>>>>>>>>>>>>>%
%======================================%

%T1>Introduction
\section{Introduction}
\label{sec:introduction}

The dynamical $p$-brane solutions in a higher-dimensional 
gravity theory were studied by \cite{Gibbons:2005rt, Chen:2005jp, 
Kodama:2005fz, Kodama:2005cz, 
Kodama:2006ay, Binetruy:2007tu, Binetruy:2008ev, Maeda:2009tq, 
Maeda:2009zi, Gibbons:2009dr, Maeda:2009ds, Uzawa:2010zza, Maeda:2010ja, 
Minamitsuji:2010fp, Maeda:2010aj, Minamitsuji:2010kb, Uzawa:2010zz, 
Nozawa:2010zg, Minamitsuji:2010uz, Maeda:2011sh, Minamitsuji:2011jt, 
Maeda:2012xb, Blaback:2012mu, Minamitsuji:2012if, Uzawa:2013koa, 
Uzawa:2013msa, Blaback:2013taa, Uzawa:2014kka, Uzawa:2014dra, Maeda:2015joa, 
Maeda:2016lqw} 
and have been widely discussed ever since. 
However, some aspects of the physical properties, 
such as supersymmetry and its breaking in the context of 
string theory, have remained slightly unclear. The motivation 
for this work is to improve this situation. For
this purpose, it is first necessary to construct supersymmetric 
brane solutions depending on the time as well as space
coordinates.

In the static background, an 
M-brane solution in the 11-dimensional supergravity 
has been constructed, and the properties have been discussed~\cite{Duff:1990xz}. In the dynamical background, it is well known that 
there are $p$-brane solutions with dynamical several $p$-brane objects in 
the expanding Universe. 
The first example was found for a D3-brane in the ten-dimensional 
type IIB string theory \cite{Gibbons:2005rt}, which was generalized for complicated field 
configurations \cite{Binetruy:2007tu, Maeda:2009zi}. There has been, however, little success 
at constructing the dynamical $p$-brane solution preserving supersymmetry, nor has there been much 
insight about what kind of geometrical structure might be expected.

The dynamical M2-brane background preserving supersymmetry is a kind 
of natural extension of the static M2-brane system, which can be described by 
an analogous Reissner-Nordstr\"{o}m solution in the 
four-dimensional Einstein-Maxwell theory. 
The existence of supersymmetry in a dynamical background 
should not come as a surprise, since several analytic solutions in string 
theories are already known \cite{Blau:2001ne, Sakaguchi:2003cy, 
Kodama:2005fz}. 
In this paper, we will find the supersymmetric 
dynamical M-brane as an exact solution of the 
supergravity field equations. What we will construct is a time-dependent 
M2-brane solution preserving supersymmetry in the 11-dimensional supergravity 
theory. Depending on which ansatz we take, we thus obtain a black hole in the 
expanding Universe. 
Although it is not necessarily easy, the supersymmetric black hole models 
governing the dynamics of the Universe 
can be constructed analytically because these 
are given by the classical solution of field equations.

Different forms of the dynamical brane solution we will be discussing have 
been obtained by \cite{Kodama:2005fz} as a supersymmetric solution in 
a ten-dimensional type IIB 
string theory and by \cite{Binetruy:2007tu, Maeda:2009zi} as a cosmological solution without 
supersymmetry for an 
11-dimensional supergravity model. A class of classical black hole solutions 
in the expanding Universe was found by \cite{Maeda:2010ja, Nozawa:2010zg}. 
Time dependent black hole solutions in lower dimensional effective field 
theories derived from string theory have been analyzed 
in~\cite{Maeda:2009zi, Maeda:2011sh}.

Although we will consider in this paper the 11-dimensional supergravities, 
there is also a ten-dimensional version of the supersymmetric dynamical 
D-brane solutions. It can be obtained by compactifying an internal space. 
In terms of the dimensional reduction of a dynamical 
M-brane background to the string in ten dimensions, 
the solution leads to the dynamical D-brane systems. 
One starts with an 11-dimensional model, but the resulting ten-dimensiomal 
model turns out to have a dynamical D-brane, as in the construction of 
\cite{Minamitsuji:2010kb}. 

This paper is organized as follows. We present an exact 
solution having a quarter of a full supersymmetry 
for a dynamical M-brane in an 11-dimensional supergravity 
and discuss how to break supersymmetries in Sec.~\ref{sec:dm}. 
In the remainder of the paper, 
we describe some applications of the result, which are   
the behavior of the geodesic, the analysis of the geometrical structure, 
and the evolution of a time-dependent black hole    
in dynamical M2-brane background. In Sec.~\ref{sec:gs}, 
we start our discussion of supersymmetric M2-brane solution by 
examining the basic features of the background geometry. 
By solving the radial null geodesic equations, 
we show that the naked strong curvature singularity appears.   
Then we investigate extensions of the solutions 
inside the horizon and discuss the smoothness at the horizon.  
In Sec.~\ref{sec:sb}, 
we present that the M2-brane background gives a black hole solution in
a time-dependent universe and discuss their implications to 
lower-dimensional effective theories. 
Section \ref{sec:discussions} contains some discussions and 
concluding remarks. 

%======================================%
%<<<<<<<<<<<<< SECTION 2 >>>>>>>>>>>>>>%
%======================================%
%T1>Dynamical M-brane backgrounds
\section{Dynamical M-brane backgrounds}
\label{sec:dm}

In this section, we will construct the exact solution to the 
field equations of an 11-dimensional supergravity corresponding 
to a dynamical M2-brane configuration. The 11-dimensional gravitino 
(Killing spinor field) equation gives the time-dependent solution 
with the particular ansatz of fields. We find that 
the supersymmetric solution depends on the null coordinate along the 
M2-brane world volume, as well as the coordinates of the transverse 
space to the M2-brane. 
%%%%%%%%%%%%%%%%%%%%%%%%%%%%%%%%%%%%%%%%%%%%%%%%
%%%%%%%%%%%%%%%%%%%%%%%%%%%%%%%%%%%%%%%%%%%%%%%%
%T2>Supersymmetry in a dynamical M2-branes
\subsection{Supersymmetry in a dynamical M2-brane}
%%%%%%%%%%%%%%%%%%%%%%%%%%%%%%%%%%%%%%%%%%%%%%%%
%%%%%%%%%%%%%%%%%%%%%%%%%%%%%%%%%%%%%%%%%%%%%%%%
We will start by making an 
ansatz for an 11-dimensional metric $g_{MN}$ and 
three-form gauge potential $A_{(3)}$. The 11-dimensional 
metric and gauge potential are assumed to be 
\Eqrsubl{s:ansatz:Eq}{
ds^2&=&A^2(x, y)\eta_{\mu\nu}(\Xsp)dx^{\mu}dx^{\nu}+B^2(x, y)
\delta_{ij}(\Ysp)dy^idy^j\,,
  \label{s:metric:Eq}\\
A_{(3)}&=& \chi C(x, y)\Omega(\Xsp)\,,
  \label{s:F:Eq}
}
where 
$\mu,\, \nu=0,\,1,\,2$, and $i,\,j=3,\,4,\,\cdots, 10$, $\chi=\pm 1$\,, and 
$\Omega(\Xsp)$ denotes the volume form of the three-dimensional Minkowski 
space~(X space). 
\,\,All components of the gravitino $\psi_M$ are zero.  
The arbitrary functions $A$, $B$, and $C$ depend on 
the M2-brane world volume coordinates $x^{\mu}$ as well as 
the radial coordinate of 
the eight-dimensional Euclidean space~(Y space)
\Eq{
r^2=\delta_{ij}y^iy^j\,.
}
Then, the metric of Y space becomes 
\Eq{
\delta_{ij}(\Ysp)dy^idy^j=dr^2+r^2u_{ab}(\Zsp)dz^adz^b\,,
  \label{s:metric-z:Eq}
}
where $u_{ab}(\Zsp)$ denotes the metric of the seven-sphere. 
As we will find that three functions $A$, $B$, and $C$ 
are reduced to one by the requirement that the metric 
and gauge field preserve supersymmetry. 
Then 
we find a 11-dimensional Killing spinor $\varepsilon$ satisfying
\Eq{
\bar{\na}_M\varepsilon=0\,. 
  \label{s:Killing:Eq}
} 
Here, $\bar{\na}_M$ is the supercovariant derivative 
appearing in the supersymmetry transformation rule 
of the gravitino
\Eq{
\bar{\na}_M=\pd_M
+\frac{1}{4}{\omega_M}^{PQ}{\Gamma}_{PQ}
+\frac{1}{12} \left({\Gamma}_M\,{\bf F}
     -3{\bf F}_M\right)
\,,
}
in terms of the 11-dimensional $\gamma$-matrices ${\Gamma}^M$  
satisfying 

\Eq{
{\Gamma}^M {\Gamma}^N + {\Gamma}^N {\Gamma}^M=2g^{MN}.
}
${\bf F}$ and ${\bf F}_M$ are defined by 
\Eqrsubl{st:def:Eq}{
{\bf F}&=&\frac{1}{4!}F_{MNPQ}{\Gamma}^{MNPQ}\,,\\
{\bf F}_M&=&
\frac{1}{3!}F_{MNPQ}{\Gamma}^{NPQ}\,,
}
and $F_{(4)}$ is the field strength defined by the 
three-form gauge potential $A_{\left(3\right)}$\,,
\Eq{
F_{(4)}=dA_{(3)}\,.
\label{def:four-form}
} 
The notation that has been used here is
\Eq{
{\Gamma}_{MN\dots P}={\Gamma}_{\left[M\right.}{\Gamma}_N
\cdots{\Gamma}_{\left.P\right]}\,.
}

For the background (\ref{s:metric:Eq}) and 
(\ref{s:metric-z:Eq}), it is convenient to introduce $\gamma_{\mu}~
(\mu=0\,,~1\,,~2)$, $\gamma_r$\,, and $\gamma_a~(a=4\,,~\cdots\,,~10)$ by
\Eq{
{\Gamma}^{\mu}=A^{-1}\gamma^{\mu}\,,~~~
{\Gamma}^r=B^{-1}\gamma^r\,,~~~
{\Gamma}^a=\frac{1}{rB}\gamma^a\,.
}
Then, $\gamma^\mu$ gives the SO(2, 1) $\gamma$-matrices, $\gamma^a$ provides 
the $\gamma$ matrices of Z, and $(\gamma^r)^2=1$.
We also define $\gamma_{(3)}$ as 
\Eq{
\gamma_{(3)}:=\gamma_0\gamma_1\gamma_2\,. 
}

We take an ansatz for the 11-dimensional metric (\ref{s:metric:Eq})
\Eq{
A=C^{1/3}\,,~~~~B=C^{-1/6}\,,~~~~C\equiv h^{-1}(x, r)\,.
  \label{s:metric2:Eq}
}
Then, in terms of these $\gamma$ matrices, the supercovariant derivative  
in the background with the metric (\ref{s:metric:Eq}), (\ref{s:metric-z:Eq}), 
and field (\ref{s:F:Eq}) is expressed as
\Eqrsubl{s:cd:Eq}{
\bar{\na}_\mu&=&\pd_\mu+\frac{1}{6}\pd_\nu\ln h {\gamma^\nu}_\mu
-\frac{1}{6}h^{-3/2}\pd_rh\gamma^\mu \gamma^r
\left(1-\chi\gamma_{(3)}\right)\,,\\
\bar{\na}_r&=&\pd_r
-\frac{1}{12}h^{-1/2}\pd_\nu h\gamma^\nu\gamma^r
+\frac{1}{6}\chi h^{-1}\pd_r h\gamma_{(3)}\,,\\
\bar{\na}_a&=&{}^\Zsp\na_a-\frac{r}{12}h^{-1/2}
\pd_\nu h\gamma^\nu \gamma_a
-\frac{r}{12}h^{-1}\pd_r h\gamma^r \gamma_a\left(1-\chi\gamma_{(3)}\right)\,,
}
where ${}^\Zsp\na_a$ is the
covariant derivative with respect to the metric $u_{ab}(\Zsp)$\,. 
The number of unbroken supersymmetries is given by the number of Killing spinor $\varepsilon$\,. 
The Killing spinor equation (\ref{s:Killing:Eq}) is automatically satisfied provided that the 
following three conditions are satisfied: 
\Eq{
\varepsilon=h^{-1/6}\varepsilon_0\,,~~~~~
\pd_\mu h\gamma^\mu\,\varepsilon=0\,,~~~~~
\left(1-\chi\gamma_{(3)}\right)\varepsilon=0\,,
  \label{s:SUSY:Eq}
}
where the sign $\chi$ comes from the ansatz of 
the three-form gauge potential (\ref{s:F:Eq}), and 
$\varepsilon_0$ 
denotes a  
constant Killing spinor. 

%%%%%%%%%%%%%%%%%%%%%%%%%%%%%%%
%%%%%%%%%%%%%%%%%%%%%%%%%%%%%%%
%T2>Einstein equations
\subsection{Einstein equations}
%%%%%%%%%%%%%%%%%%%%%%%%%%%%%%%
%%%%%%%%%%%%%%%%%%%%%%%%%%%%%%%
Now we determine a form of the function $h(x, r)$ in an 
11-dimensional supergravity theory which is 
composed of the metric $g_{MN}$ and the four-form field strength $F_{(4)}$. 
The action in 11 dimensions is 
given by 
\Eqr{
S=\frac{1}{2\kappa^2}\int \left[R\,\ast{\bf 1}
 -\frac{1}{2\cdot 4!}
 \ast F_{(4)}\,\wedge\, 
 F_{\left(4\right)} \right]
 -\frac{1}{12\kappa^2}\int A_{(3)}\,\wedge\, 
 F_{\left(4\right)}\wedge\, 
 F_{\left(4\right)}\,,
\label{s:a:Eq}
}
where $R$ denotes the Ricci scalar 
with respect to the 11-dimensional metric $g_{MN}$\,, 
$\kappa^2$ is the 11-dimensional gravitational constant, 
$\ast$ denotes the Hodge operator in the 11-dimensional spacetime, and 
$F_{\left(4\right)}$ is the four-form field strength 
defined by (\ref{def:four-form}), respectively. 

Let us first consider the gauge field equation
\Eq{
d\left(\ast F_{(4)}\right)+\frac{1}{2}F_{\left(4\right)}\wedge\, 
 F_{\left(4\right)}=0\,.
\label{s:gauge:Eq}
}
Using the ansatz of fields (\ref{s:metric:Eq})\,, 
(\ref{s:metric2:Eq}), the above equation is reduced to   
\Eq{
\pd_\mu \pd_r h=0\,,~~~~~~\left(\pd_r^2+\frac{7}{r}\pd_r\right)h=0\,.
\label{s:gauge2:Eq}
 }
From Eq.~(\ref{s:gauge2:Eq})\,, 
the function $h$ and the field equation can be expressed as 
\Eq{
h(x, r)=h_0(x)+h_1(r)\,,~~~~~~\left(\pd_r^2+\frac{7}{r}\pd_r\right)h_1=0\,.
  \label{s:h:Eq}  
}
Then, imposing the boundary condition that the 11-dimensional 
metric is asymptotically vacuum spacetime, we find 
\Eq{
h_1(r)=\tilde{c}+\frac{M}{r^6}\,,
    \label{s:h1:Eq}
}
where $\tilde{c}$ is constant. 

Next we show that Eq.~(\ref{s:h1:Eq}) is consistent with the Einstein 
equations and derive the equation for the function $h_0$.

The Einstein equations are given by 
\Eq{
R_{MN}=\frac{1}{2\cdot 4!}
\left[4F_{MABC} {F_N}^{ABC}
-\frac{1}{3}\,g_{MN}\,F^2_{\left(4\right)}\right].
   \label{s:Einstein:Eq}
}
Using the assumptions (\ref{s:ansatz:Eq}) and 
(\ref{s:metric2:Eq}), Einstein equations become 
\Eqrsubl{s:cEinstein:Eq}{
&&-h^{-1}\pd_{\mu}\pd_{\nu} h +\frac{1}{3}h^{-1} 
\eta_{\mu\nu}\left[\triangle_{\Xsp}h + h^{-1}
\left(\pd_r^2+\frac{7}{r}\pd_r\right)h\right]=0\,,
 \label{s:cEinstein-mu:Eq}\\
&& 
\triangle_{\Xsp} h
 +h^{-1}\left(\pd_r^2+\frac{7}{r}\pd_r\right)h=0 %}
\,,
 \label{s:cEinstein-rr:Eq}\\
&&R_{ab}(\Zsp)-6u_{ab}(\Zsp)-\frac{1}{6} r^2u_{ab}(\Zsp)
\left[\triangle_{\Xsp} h
 +h^{-1}\left(\pd_r^2+\frac{7}{r}\pd_r\right)h \right]=0\,,
 \label{s:cEinstein-ab:Eq}\\
&&\pd_{\mu}\pd_r h=0,
 \label{s:cEinstein-mi:Eq}
}
where $\triangle_{\Xsp}$ is the Laplace operator on the space of 
${\rm \Xsp}$\,, and $R_{ab}(\Zsp)$ is the Ricci tensor 
of the metric $u_{ab}(\Zsp)$\,.
From Eq.~(\ref{s:cEinstein-mi:Eq}), 
the function $h$ must be in the form
\Eq{
h(x, r)= h_0(x)+h_1(r).
  \label{eq:sec2:form of warp factor}
}
With this form of $h$, 
the Einstein equations reduce to  
\Eqrsubl{s:cEinstein3:Eq}{
&&R_{ab}(\Zsp)=6u_{ab}(\Zsp),\\
   \label{s:Ricci-Z:Eq} 
&&\pd_{\mu}\pd_{\nu}h_0=0\,.
   \label{s:h-sol:Eq} 
 }
In this case, the first equation is automatically satisfied,  %} 
and the solution for 
$h$ can be written explicitly as
\Eq{ 
h(x, r)=c_{\mu}x^{\mu}+\bar{c}
+\frac{M}{r^6},
 \label{s:h-sol2:Eq}
}
where $c_{\mu}$, $\bar{c}$, and $M$ are constant parameters. 
As seen from supersymmetric equations (\ref{s:SUSY:Eq}), the parameters 
$c_\mu$ have to obey the relation (\ref{s:SUSY:Eq}), which is 
given by $c_\mu \gamma^\mu\varepsilon_0=0$\,. 
So, without loss of generality, we shall impose that $\bar{c}=0$ 
and $c_\mu x^\mu=c(t-x)/\sqrt{2}$, where $c$ is a constant.

%%%%%%%%%%%%%%%%%%%%%%%%%%%%%%%%%%%%%%%%%%%%%%%%
%%%%%%%%%%%%%%%%%%%%%%%%%%%%%%%%%%%%%%%%%%%%%%%%
%%%%%%%%%%%%%%%%%%%%%%%%%%%%%%%%%%%%%%%%%%%%%%%%
%T2>Number of SUSY
\subsection{Number of supersymmetry and supersymmetry breaking}
\label{sub:n}
In this section, we count the number of preserving supersymmetry 
in the dynamical M2-brane background. 
An unbroken supersymmetry with respect to each Killing spinor 
$\varepsilon$ has to obey the integrability condition
\Eq{
[\bar{\nabla}_M ,~ \bar{\nabla}_N] \varepsilon =0\,.
   \label{n:int:Eq}
}
From the relation 
\Eq{
\nabla_M=\pd_M
+\frac{1}{4}{\omega_M}^{PQ}{\Gamma}_{PQ}\,,\quad
\left[\nabla_M ,~ \nabla_N\right]
      =\frac{1}{4} R_{MNPQ} \Gamma^{PQ}\,,
   \label{eq:sub-condition}
}
the commutator of the covariant derivatives in the integrability  
condition (\ref{n:int:Eq}) becomes
\begin{eqnarray}
[\bar{\nabla}_M ,~ \bar{\nabla}_{N}]
   &=&\frac{1}{4} R_{MNPQ} \Gamma^{PQ}
      +\frac{1}{6} \left(\nabla_{[M}\,\Gamma_{N]}\,{\bf F}
      -3\nabla_{[M}\,{\bf F}_{N]}\right)\nn\\
      &&+\frac{1}{144} \left[ 
        \left({\Gamma}_M\,{\bf F}-3{\bf F}_M\right)\,,~
        \left({\Gamma}_N\,{\bf F}-3{\bf F}_N\right)     
        \right]. 
   \label{eq:integ-condition1}
\end{eqnarray}
 
In terms of the condition, we count how many supersymmetries exist. 
We first briefly review the results for the 
well-known case of the 11-dimensional static background 
\cite{Nahm:1977tg, Cremmer:1978km, Freund:1980xh, Duff:1990xz}. 
For the case in which $h=\const $ or $h=M/r^6$ in the 
11-dimensional metric (\ref{s:metric2:Eq}),  
the number of supersymmetries reduce to 
the number of solutions to the spinor equation,  
$\bar{\nabla}_a\varepsilon=0$\,. 
In particular, for the 11-dimensional Minkowski 
spacetime \cite{Nahm:1977tg, Cremmer:1978km} 
and for $\AdS^4\times {\rm S}^7$, $\AdS^7\times 
{\rm S}^4$ \cite{Freund:1980xh}, 
the background has the full supersymmetry.

Next, we consider the static M2-brane background with 
$h(r)=\tilde{c}+M/r^6$ ($\tilde{c} M\not=0$) \cite{Duff:1990xz}\,,
where $\tilde{c}$ is constant. 
Then, the $\mu r$ component of the integrability condition 
gives 
\Eq{
0=[\bar{\nabla}_{\mu} ,~ \bar{\nabla}_{r}] \varepsilon 
 =-h^{-1/3}\,\frac{d^2}{dr^2} \left(h^{-1/6}\right)\,\gamma_{\mu}\,\gamma^{r}
\,\left(1-\chi\gamma_{(3)}\right)\varepsilon\,.
    \label{n:s-int:Eq}
}
Hence, $\varepsilon$ have to obey 
\Eq{
\left(1-\chi\gamma_{(3)}\right)\varepsilon=0\,.
    \label{eq:static-killing}
}
Since we can show that this condition and \eqref{n:s-int:Eq} 
are the only nontrivial integrability conditions, 
one half of the supersymmetries in the 
case $\tilde{c} M\ne 0$ is broken 
in M2-brane background \cite{Duff:1990xz}.

Now, we consider the background with $\pd_{\mu} h\ne 0$\,. 
The $[\mu,\nu]$ components of the integrability condition give 
\Eq{
0=\xi^{\mu} \zeta^{\nu} [\bar{\nabla}_{\mu} ,~ \bar{\nabla}_{\nu}] 
  \varepsilon
 =-\frac{\eta^{\mu\nu}\pd_\mu h_0\pd_\nu h_0}
   {18 h^{2}} \xi_{\rho}\gamma^{\rho} \zeta_{\sigma} 
 \gamma^{\sigma} \varepsilon\,,
   \label{eq:susy-condition}
}
where $\xi^{\mu}$ and $\zeta^{\nu}$ are linearly independent vectors 
satisfying the conditions\,, $\xi^{\mu} \pd_{\mu} h=\zeta^{\mu} \pd_{\mu} 
h=0$\,, and we assume that the function $h(x, r)$ obeys 
\Eq{
h(x, r)=h_0(x)+h_1(r)\,,~~~\pd_\mu\pd_\nu h_0=0\,. 
}  
Hence, it follows that if $c_\mu=\pd_\mu h_0$ is not null, 
there exists only a trivial solution to the Killing spinor equation, 
and the supersymmetry is completely broken. On the other hand, when $c_\mu$ 
is a null vector, the Killing spinor equation leads to (\ref{s:SUSY:Eq}). 
For the case  
\Eq{ 
h(x, r)=c_\mu x^\mu+\tilde{c}+\frac{M}{r^6}\,,
 \label{n:h:Eq}
}
one quarter of the possible rigid supersymmetries 
in the maximal case survives.

Here, we check the degree of supersymmetry for the case of $M=0$. 
An important simplification occurs if we consider the following special 
case of vanishing M2-brane charge:
\Eqrsubl{n:pp:Eq}{
ds^2&=&h^{-2/3}(u)\left[-2dudv+\left(dy\right)^2
\right]+h^{1/3}(u)\,\delta_{mn}dz^mdz^n\,,\\
h(u)&=&c\,u\,,~~~~u=\frac{1}{\sqrt{2}}(t-x)\,,~~~~
v=\frac{1}{\sqrt{2}}(t+x)\,,
} 
from the dynamical M2-brane to the plane wave background. Here, $c$ is 
constant, and $\delta_{mn}$\,,  
$z^m$ denote the metric, coordinates of eight-dimenional Euclidean space, 
respectively. The required change of coordinates is 
$(u, v, z^m)\rightarrow(\bar{u}, \bar{v}, \bar{z}^m)$\,, where
\Eq{
u=\frac{1}{c}\left(\frac{\bar{u}}{\bar{u}_0}\right)^3\,,~~~~~
v=\bar{v}+f(\bar{u})\,\delta_{mn}\bar{z}^m\bar{z}^n\,,~~~~~
z^m=h^{-1/6}(\bar{u})\bar{z}^m\,,
}
which leads to the 
plane wave metric \cite{Papadopoulos:2002bg},
\Eq{
ds^2=-2d\bar{u}d\bar{v}
+\left(\frac{\bar{u}}{\bar{u}_0}\right)^{-2}
\left[-\frac{c^2}{36}\,\delta_{mn}\bar{z}^m\bar{z}^n\left(d\bar{u}\right)^2
+\left(dy\right)^2\right]+\delta_{mn}d\bar{z}^md\bar{z}^n\,.
\label{n:pp2:Eq}
}
Here, we used 
\Eq{
\bar{u}_0=\frac{3}{c}\,,~~~~~
f(\bar{u})=-\frac{c}{12}\left(\frac{\bar{u}}{\bar{u}_0}\right)^{-1}\,.
}
Setting $M=0$ in the solution (\ref{s:h-sol2:Eq}), 
the integrability condition 
reduces to $c_{\mu}\,\gamma^{\mu} \varepsilon=0$\,. Then, 
the dynamical M2-brane solution with $c_\mu\ne 0$\,,
 preserves a half of the maximal 
supersymmetries. Since the number of unbroken spacetime supersymmetries
in the present background must be a half of the full supersymmetries, 
as in a generic plane wave, our solution is consistent with past results  
 \cite{Papadopoulos:2002bg}.  

%%%%%%%%%%%%%%%%%%%%%%%%%%%%%%%%%%%%%%%%%%%%%%%%
%%%%%%%%%%%%%%%%%%%%%%%%%%%%%%%%%%%%%%%%%%%%%%%%
%%%%%%%%%%%%%%%%%%%%%%%%%%%%%%%%%%%%%%%%%%%%%%%%

Next we comment on the degree of the supersymmetry breaking for the 
dynamical M2-brane background. 
The measure of the supersymmetry breaking for the dynamical background is 
obtained from the consistency condition. The mass scale corresponds to 
$h^{-2}\eta^{\mu\nu}\pd_\mu h\pd_\nu h$, 
which could be identified with a kind of induced effective mass scale for 
the spinor field. The divergence at $h=0$ 
means that the degree of the supersymmetry breaking increases as the 
background approaches the curvature singularity. On the other hand, 
the supersymmetry breaking becomes negligible near the M2-brane region 
$r\rightarrow 0$\,, 
as $h$ diverges there. 

Let us consider the relation between the dynamics of the background 
and supersymmetry breaking in more detail. 
Introducing a new time coordinate $\tau$\,, which is defined by 
$\tau/\tau_0=(c_0\,t)^{2/3}$\,, with constant $\tau_0=(3/2c_0)$\,, 
 we find the 11-dimensional metric (\ref{s:metric:Eq}) as 
\Eqr{
ds^2&=&-\left[1+\left(\frac{\tau}{\tau_0}\right)^{-3/2}
\left(c_ix^i+\frac{M}{r^6}\right)\right]^{-2/3}
\left[-d\tau^2+\left(\frac{\tau}{\tau_0}\right)^{-1}
\delta_{ij}dx^i dx^j\right]\nn\\
&&+\left[1+\left(\frac{\tau}{\tau_0}\right)^{-3/2}
\left(c_ix^i+\frac{M}{r^6}\right)\right]^{1/3}
\left(\frac{\tau}{\tau_0}\right)^{1/2}
\left[dr^2+r^2d\Omega_{(7)}^2\right], 
}
where $x^i~(i=1, 2)$ denotes the space coordinates of the 
world volume spacetime, and the metric $\delta_{ij}$ is the 
spatial part of the three-dimensional Minkowski metric 
$\eta_{\mu\nu}$\,. 
When we set $c_1=c_2=0$\,, the spacetime is an isotropic and 
homogeneous universe with respect to the world volume coordinates, 
whose supersymmetry is completely broken. 
On the other hand, the 11-dimensional spacetime becomes 
inhomogeneous and preserves supersymmetry 
if parameters $c_\mu$ satisfy $c_\mu c^\mu=0$\,, and 
$c_\mu\gamma^\mu\varepsilon=0$\,. 
Thus, in the limit when the terms $c_ix^i$ are negligible, 
which is realized in the limit $(\tau/\tau_0)\rightarrow\infty$\,,  
for small $r$\,, 
we find an 11-dimensional universe without supersymmetry. 
For concreteness, we discuss the dynamics in the region where 
the term $c_i x^i$ in the function 
$h(\tau, x, r)$ is smaller compared to the contribution 
of the M2-brane charge $M/r^6$\,. 
In the case of $(\tau/\tau_0)>0$\,, we have found that the domains 
near the M2-brane has the supersymmetry. 
As the time increases,
 the background satisfies $({\tau}/{\tau_0})^{3/2}\gg c_ix^i$\,, 
Then, we find
\Eq{
1+\left(\frac{\tau}{\tau_0}\right)^{-3/2}
\left(c_ix^i+\frac{M}{r^6}\right)
\rightarrow 1+\left(\frac{\tau}{\tau_0}\right)^{-3/2}
\frac{M}{r^6}\,.  
}
The contribution of the term $c_i x^i$ in the function 
$h(\tau, x, r)$ eventually 
becomes negligible in the 11-dimensional metric such that 
supersymmetries are completely broken, which is guaranteed by the 
region $c_i x^i\ll M/r^6$\,. 
Then, the dynamical M2-brane solution also behaves as a 
nonsupersymmetric cosmological solution in the asymptotic future. 

Finally, we also comment about a relation between the M2-brane or 
black hole and plane wave background. 
Now, we set
\Eq{
h(t, x, r)=\frac{c}{\sqrt{2}}(t-x)+\frac{M}{r^6}\,.
}
In the limit when the term $M/r^6$
is negligible, corresponding to the far region from the M2-brane, 
the background changes from the above description to 
a time-dependent plane wave background (\ref{n:pp:Eq}). 
Hence, the supersymmetry will enhance from 
one quarter to a half of the possible rigid supersymmetries  
in the maximal case when one moves in the 
transverse space in such a way that $({\tau}/{\tau_0})^{-3/2}c_i x^i$ 
remains approximately constant. 
Although the solution itself is by no means realistic, 
its interesting behavior suggests an enhancement of the 
supersymmetry, or a possibility that the Universe with a quarter of 
the preserved original supersymmetry began to evolve toward a universe  
without supersymmetry.

%======================================%
%<<<<<<<<<<<<< SECTION 3 >>>>>>>>>>>>>>%
%======================================%
%T1>Geometry of the supersymmetric dynamical M2-brane solution
\section{Geometry of the supersymmetric dynamical M2-brane solution}
\label{sec:gs}
As one may expect from the dynamical M2-brane solution, 
the spacetime with (\ref{s:h-sol2:Eq}) has curvature singularity. 
For a fixed $x$, the spacetime asymptotically approaches the  
anisotropic solution at a large $r$, while the metric becomes approximately 
AdS${}_4\times$S${}^7$ near the M2-brane region 
(at $r\rightarrow 0$)\,, as we will show it in the 
Sec.~\ref{sec:nu}\,. 
Now we investigate the geometric feature near the curvature singularity and 
discuss the smoothness at the horizon. 

%%%%%%%%%%%%%%%%%%%%%%%%%%%%%%%%%%%%%
%%%%%%%%%%%%%%%%%%%%%%%%%%%%%%%%%%%%%
%T2>Property of the solution
\subsection{Property of the solution}
\label{sec:p}
%%%%%%%%%%%%%%%%%%%%%%%%%%%%%%%%%%%%%
%%%%%%%%%%%%%%%%%%%%%%%%%%%%%%%%%%%%%
We consider the following time dependent M2-brane 
solution with the 11-dimensional metric 
\Eqrsubl{p:bg:Eq}{
&&ds^2=h^{-2/3}(u, r)\left(-2du\,dv+dy^2\right)
+h^{1/3}(u, r)\left[dr^2+r^2d\Omega_{(7)}^2\right]\,,
 \label{p:metric:Eq}\\
&&u=\frac{1}{\sqrt{2}}\left(t-x\right)\,,~~~~~~
v=\frac{1}{\sqrt{2}}\left(t+x\right)\,,\\
&&h(u, r)=h_0(u)+h_1(r),~~~~~h_0(u)=c u
\,,~~~~~
h_1(r)=\frac{M}{r^6}\,, 
  \label{p:h:Eq}
}
where $c$, $M$ are constants.
Since the function $h_1(r)$ dominates near $r\rightarrow 0$\,, the background 
geometry describes the extremal Reissner-Nordstr\"{o}m solution with 
an infinite 
throat. The geometry of the dynamical M2-brane is not asymptotically flat 
while the extremal Reissner-Nordstr\"{o}m solution gives the 
asymptotically Minkowski spacetime. 
Near the M2-brane, the metric becomes AdS${}_4\times$ S${}^7$\,,
\Eqr{
ds^2&\approx& \frac{r^4}{M^{2/3}}\left(-2du\,dv+dy^2\right)
+\frac{M^{1/3}}{r^2}dr^2+M^{1/3}d\Omega_{(7)}^2\,,\nn\\
&=&\frac{M^{1/3}}{4w^2}(-2du\, dv+dy^2+dw^2)+M^{1/3}d\Omega_{(7)}^2\,, 
\qquad w:=\frac{M^\frac{1}{2}}{2r^2}\,,
}
where $d\Omega_{(7)}^2$ is the line element of the seven-sphere. 

Since the square of the four-form field strength diverges 
at the zeros of the function $h(u, r)=0$\,, 
\Eq{
F^2_{(4)}=-4!\,h^{-7/3}\left(\pd_r h\right)^2\,,
}
the curvature of the metric (\ref{p:bg:Eq}) can be singular 
at $h(u, r)=0$\,.

Now we discuss the cosmological evolution of the spatial geometry 
in the region $h>0$
and assume $c<0$, in the function $h(u, r)$. 
For $u<0$\,, the function $h$ is positive everywhere and the spatial surfaces 
are nonsingular unless we treat the negative charge of the M2-brane $M<0$\,. 
They are asymptotically anisotropic spacetime for 
a fixed $x$ coordinate. 
The spatial metric is still regular for $u=0$ besides the region 
$r\rightarrow\infty$\,. As time increases slightly,  
a singularity appears at $r=\infty$ and moves in from spatial infinity. 
As $u$ evolves further, the singularity 
eventually wraps the horizon completely.   

%%%%%%%%%%%%%%%%%%%%%%%%%%%%%%%%
%%%%%%%%%%%%%%%%%%%%%%%%%%%%%%%%
%T2>Geodesic motion
\subsection{Geodesic motion}
\label{sec:lh}
%%%%%%%%%%%%%%%%%%%%%%%%%%%%%%%%
%%%%%%%%%%%%%%%%%%%%%%%%%%%%%%%%

We start by solving radial null geodesic equations for the affine 
parameter $s$ on the background (\ref{p:metric:Eq}). 
As found in~\cite{Maeda:2015joa}, the geodesic equations are 
\begin{align}
& \frac{du}{ds}=fh^{2/3}\,,~~~~\frac{dv}{ds}=\frac{h^{1/3}}{2f}\left(\frac{dr}{ds}\right)^2, \nonumber \\
& \frac{d^2r}{ds^2}=-\frac{c}{3}fh^{-1/3}\frac{dr}{ds}-\frac{M}{r^7}h^{-1}
\left(\frac{dr}{ds}\right)^2\,,
   \label{lh:ge:Eq}
\end{align}
where $f$ is a constant. 

%%%%%%%%%%%%%%%%%%%%%%%%%%%%%%%%
%%%%%%%%%%%%%%%%%%%%%%%%%%%%%%%%
%T3>Geodesic motion near M2-brane
\subsubsection{Geodesic motion near the M2-brane}
\label{sec:gmb}
%%%%%%%%%%%%%%%%%%%%%%%%%%%%%%%%
%%%%%%%%%%%%%%%%%%%%%%%%%%%%%%%%

Near the M2-brane, the null geodesic solution of Eq.~(\ref{lh:ge:Eq}) is found analytically. 
Let us assume that $|u|\ll r^{-6}$ in the limit of $r\rightarrow 0$\,. 
Then, the function $h$ takes the simple form 
\Eq{
h\rightarrow \frac{M}{r^6}\,. 
}
In this approximation, 
the asymptotic solution is given by    
\begin{align}
u\sim (s_0-s)^{-1}, \qquad r\sim \sqrt{s_0-s}\,, 
\end{align}
near $r=0$, where $s_0$ is a positive constant. Note that 
the assumption $|u|\ll r^{-6}$ is satisfied in this asymptotic solution. 
Obviously, $u$ becomes infinite as we approach the location of the 
M2-brane $r\rightarrow 0$~($s\to s_0$).

%%%%%%%%%%%%%%%%%%%%%%%%%%%%%%%%
%%%%%%%%%%%%%%%%%%%%%%%%%%%%%%%%
%T3>Geodesic motion near the timelike singularity
\subsubsection{Geodesic motion near the timelike singularity}
\label{sec:gms}
%%%%%%%%%%%%%%%%%%%%%%%%%%%%%%%%
%%%%%%%%%%%%%%%%%%%%%%%%%%%%%%%%
We now discuss the radial null geodesic near the timelike singularity. For the 
supersymmetric M2-brane background (\ref{p:bg:Eq}), 
$h=0$ hypersurface corresponds to a 
timelike curvature singularity \cite{Maeda:2015joa} 
because $g^{MN}\ell_M\ell_N>0$ for $\ell_M=\nabla_M h$ near the singularity.  
Let us then consider the past directed null geodesics which 
can hit the curvature singularity within a finite affine parameter length. 
Now we set that as $h\rightarrow 0$\,, 
\Eq{
h(s)=\left(s_0-s\right)^\alpha\,,~~~~~r(s)\simeq r_0+r_1\left(s_0-s\right)^
\beta\,,
  \label{gms:hr:Eq}
}
where $s_0$ denotes the value of $s$ at singularity, 
$\alpha~(>0)$, $\beta$, and $r_1$ are constants determined later. 
Near the singularity, the geodesic equations~(\ref{lh:ge:Eq}) become
\Eqrsubl{gms:ge:Eq}{
\frac{du}{ds}&=&f\left(s_0-s\right)^{2\alpha/3}\,, ~~~~~~
\frac{dv}{ds}=\frac{1}{2f}\left(s_0-s\right)^{\alpha/3}\left(\frac{dr}{ds}
\right)^2
\,,
   \label{gms:ge-u:Eq}\\
\frac{d^2r}{ds^2}&=&-\frac{c}{3}fh^{-1/3}\frac{dr}{ds}-\frac{M}{r_0^7}h^{-1}
\left(\frac{dr}{ds}\right)^2\simeq -\frac{M}{r_0^7}\left(s_0-s\right)^{-\alpha}
\left(\frac{dr}{ds}\right)^2\,.  
}
Here, in the second line, we assumed that the second term in the 
r.~h.~s. is dominant.   
Substituting Eq.~(\ref{gms:hr:Eq}) into Eq.~(\ref{gms:ge:Eq}), we find
\begin{align}
\beta=\alpha\,,~~~~~~r_1=-\frac{r_0^7\left(\alpha-1\right)}{\alpha M}\,.
\label{alpha-beta_relation}
\end{align}
From the Eq.~(\ref{gms:ge-u:Eq}), the form of $u$ is given by 
\Eq{
u(s)=u_0-\frac{3f}{2\alpha+3}\left(s_0-s\right)^{1+\frac{2\alpha}{3}}\,. 
}
For $s\rightarrow s_0$\,, it follows that $u\rightarrow u_0$\,, 
and $r\rightarrow r_0$\,, Then 
expanding $h$ in Eq.~(\ref{p:h:Eq}) around $s=s_0$\,, 
we have 
\Eq{
h(s)=-\frac{3cf}{2\alpha+3}\left(s_0-s\right)^{1+\frac{2\alpha}{3}}
-\frac{6Mr_1}{r_0^7}\left(s_0-s\right)^{\alpha}\,,
   \label{gms:h:Eq}
}
where we have used $h(s_0)=0$\,. 
From the  
Eqs.~(\ref{gms:hr:Eq}), (\ref{alpha-beta_relation}), and (\ref{gms:h:Eq}), the 
constant $\alpha$ becomes $\alpha=6/5$\,. Note that this coefficient is consistent with 
the assumption that $|h^{-1/3}dr/ds|\ll |h^{-1}(dr/ds)^2|$\,.  

We now turn our attention to calculate a geometrical 
quantity in a 
parallelly propagated frame 
along the null geodesic,
\Eq{
\Gamma\equiv C_{MPNQ}{E_2}^M{E_2}^Nk^Pk^Q,
}
where $C_{MNPQ}$ is the Weyl tensor, 
$k^M$ denotes the tangent vector of null geodesic, and ${E_2}^M$ is a parallelly propagated 
spacelike unit vector orthogonal to $k^M$. These are defined by 
\Eq{
k=\frac{du}{ds}\partial_u+\frac{dv}{ds}\partial_v+\frac{dr}{ds}\partial_r\,,~~~~~
{E_2}=h^{1/3}\partial_y\,.
}
In terms of the metric (\ref{p:bg:Eq}) and 11-dimensional 
null vectors (\ref{gms:ge:Eq}) with $\alpha=6/5$\,, we find 
\Eqr{
\Gamma 
\sim\left(s_0-s\right)^{-2}\,.
   \label{gms:weyl:Eq}
} 
The shear $\sigma$ and 
the expansion rate $d\theta/ds$ of the congruence along 
the null vector $k^M$ 
diverge near the singularity as 
\Eq{
\sigma \sim \int^s \Gamma ds\sim \left(s_0-s\right)^{-1}\,,~~~~
\frac{d\theta}{ds}\sim -\sigma^2\sim -\left(s_0-s\right)^{-2}\,.
}
Then, we obtain 
\Eq{
\int^s\theta\,ds=\int^s \left(\frac{d}{ds}\ln A\right)
\,ds \sim \ln \left(s_0-s\right)\,, 
   \label{gms:ex:Eq}
}
where $A$ is the volume element of the null geodesic congruence. 
This implies that the timelike singularity is a strong type of curvature singularity~\cite{ClarkeKrolak1986}, as 
the volume element of any congruence along the radial null geodesic vanishes 
there. 
%%%%%%%%%%%%%%%%%%%%%%%%%%%%%%%%%%%%%%%%%%%%%%%%%%%%
%%%%%%%%%%%%%%%%%%%%%%%%%%%%%%%%%%%%%%%%%%%%%%%%%%%%
%T2>Analytic extension across the event horizon
\subsection{Analytic extension across the event horizon}
\label{sec:nu}
%%%%%%%%%%%%%%%%%%%%%%%%%%%%%%%%%%%%%%%%%%%%%%%%%%%%
%%%%%%%%%%%%%%%%%%%%%%%%%%%%%%%%%%%%%%%%%%%%%%%%%%%%

As shown in the previous section, there are null geodesics which 
terminate a coordinate singularity, $r=0$, $t=\infty$ in the metric~(\ref{p:metric:Eq}) 
within a finite affine parameter distance. Here, we consider an analytic extension across 
the $(r=0,\,t=\infty)$ surface and show that this surface corresponds to a regular null 
hypersurface~(horizon) generated by a null Killing vector field. 

In the $c=0$ case, the metric is static and $r=0$; the
 $t=\infty$ surface corresponds to a 
Poincare horizon in AdS${}_4\times$ S${}^7$. Thus, the near horizon geometry is clearly 
regular, and the regular metric in AdS${}_4$ part is given by  
\Eq{
\label{AdS_global}
ds_{\rm AdS_4}^2\simeq -\cosh^2\rho d\tau^2+d\rho^2+\sinh^2\rho d\Omega_{(2)}^2
\,,}
by adapting a global coordinate 
system defined by    
\Eqr{
\label{global_coordinate}
&& w=\frac{1}{\cosh \rho \cos\tau+\sinh\rho\sin\theta\sin\varphi}, \nonumber \\
&& t=\frac{\cosh\rho \sin\tau}{\cosh \rho \cos\tau+\sinh\rho\sin\theta\sin\varphi}, \nonumber \\
&& x=\frac{\sinh\rho \cos\theta}{\cosh \rho \cos\tau+\sinh\rho\sin\theta\sin\varphi}, \nonumber \\
&& y=\frac{\sinh\rho \sin\theta\cos\varphi}{\cosh \rho \cos\tau+\sinh\rho\sin\theta\sin\varphi}. 
}
So, we expect that this coordinate system also works even in the $c\neq 0$. For simplicity, 
we consider the case that $c$ becomes small. 
Then, expanding the function $h$ with respect to the parameter $c$\,, and transforming 
the metric~(\ref{p:metric:Eq}) in terms of the global coordinate~(\ref{global_coordinate}), we obtain 
\Eq{
ds^2=\frac{M^\frac{1}{3}}{4}ds_{\rm AdS_4}^2+ch_{AB}dx^Adx^B
+M^\frac{1}{3}\left(1+\frac{c M^\frac{1}{2} u}{24w^3}\right)
d\Omega_{(7)}+O(c^2)\,,
 \label{nu:metric2:Eq}
}
where $ds_{\rm AdS_4}^2$ is the AdS${}_4$ spacetime in global coordinate~(\ref{AdS_global}), 
and $h_{AB}~(A, B=0,\cdots\,, 3)$ denotes the four-dimensional metric 
which describes the deviation from the AdS${}_4$ geometry 
in terms of   
global coordinates. The metric $h_{AB}$ is a complicated function of the global coordinate 
$(\tau,\,\rho,\,\theta,\,\phi)$, but each component is regular everywhere. 
So, the $r=0~(w=\infty)$ 
surface is regular, up to $O(c)$. One can check that the $r=0~(w=\infty)$ 
surface is a null hypersurface since 
\begin{align}
g^{AB}(d\xi)_A(d\xi)_B\biggr{|}_{\xi=0}=0, \qquad \xi:=1/w=\cosh \rho \cos\tau+\sinh\rho\sin\theta\sin\varphi, 
\end{align} 
up to $O(c)$. 

Next, we consider two vectors $N=\pd_t$, $X=\pd_x$ near the horizon.   
In terms of the global coordinate (\ref{global_coordinate}), we obtain 
\Eqrsubl{nu:vec:Eq}{
N&=&\frac{\cosh\rho+\sinh\rho\cos\tau\sin\theta\sin\phi}{\cosh\rho}\pd_\tau
+\sin\theta\sin\tau\sin\phi\pd_\rho\nn\\
&&+\frac{\cosh\rho\sin\tau\cos\theta\sin\phi}{\sinh\rho}\pd_\theta
+\frac{\cosh\rho\sin\tau\cos\phi}{\sinh\rho\sin\theta}\pd_\phi\,,\\
X&=&-\cos\theta\sin\tau\tanh\rho\pd_\tau+\cos\theta\cos\tau\pd_\rho\nn\\
&&-\left(\frac{\cosh\rho\cos\tau\sin\theta}{\sinh\rho}
+\sin\phi\right)\pd_\theta
-\frac{\cos\theta\cos\phi}{\sin\theta}\pd_\phi\,.
}

Since the vector $N$ is proportional to $X$ and $g(N,\,N)=g(X,\,X)=0$ 
on the null hypersurface $\xi=0$\,, these vectors become null and degenerate 
on the horizon. So,    
\Eq{
\partial_v:=\frac{1}{\sqrt{2}}\left(N+X\right)\,, 
}
is also null on the horizon~($r=0$). Thus, the null Killing vector field $\partial_v$ is also the generator of the horizon, 
even though the bulk metric is asymptotically 
anisotropic geometry 
at constant $x$ coordinate.

%======================================%
%<<<<<<<<<<<<< SECTION 4 >>>>>>>>>>>>>>%
%======================================%
%T1>Supersymmetric black hole in an expanding universe
\section{Supersymmetric black hole in an expanding universe}
  \label{sec:sb}

The static M2-brane system
describes the microstate of a black hole \cite{Stelle:1998xg}. Then, 
it may be natural to apply the present solutions
to a time-dependent spacetime with a black hole. 
As we have presented in the previous section, there is a null Killing vector 
at the horizon where the M2-brane is located. 
In the limit $r\rightarrow 0$\,, the background geometry thus 
becomes AdS${}_4\times$S${}^7$\,. In this section, we discuss  
the dynamics of a black hole, which is so called the "black M2-brane" 
\cite{Stelle:1998xg, Horowitz:1991cd, Duff:1993ye, Duff:1996hp}, 
 in the expanding Universe on the basis of the results we have obtained 
in the previous section. 

%%%%%%%%%%%%%%%%%%%%%%%%%%%%%%%%%%%%%%%%%%%%%%%%%%%%
%%%%%%%%%%%%%%%%%%%%%%%%%%%%%%%%%%%%%%%%%%%%%%%%%%%%
%T2>Black hole in an 11-dimensional background
\subsection{Black hole in an 11-dimensional background}
%%%%%%%%%%%%%%%%%%%%%%%%%%%%%%%%%%%%%%%%%%%%%%%%%%%%
%%%%%%%%%%%%%%%%%%%%%%%%%%%%%%%%%%%%%%%%%%%%%%%%%%%%
Here, we give an explicit example of a black hole in the dynamical 
M2-brane system. The 11-dimensional metric of the 
supersymmetric M2-brane depends on time,
\Eq{
ds^2=h^{-2/3}(u, r)\left[-2dudv+\left(dy\right)^2+h(u, r)\left(dr^2
+r^2d\Omega_{(7)}\right)\right],
}
where
\Eq{
h(u, r)=cu+{M\over r^6}\,, 
}
with constants $c$ and $M$\,. If we introduce a new coordinate 
$\bar{u}$\,, this metric is rewritten as
\Eq{
ds^2=H^{-2/3}(\bar{u}, r)\left[-2d\bar{u}dv+a_{\rm M2}^{-\frac{4}{3}}
(\bar{u})\left(dy\right)^2
+a_{\rm M2}^2(\bar{u})H(\bar{u}, r)\left(dr^2
+r^2d\Omega_{(7)}\right)\right],
}
where $\bar{M}(\bar{u})$\,, $a_{\rm M2}(\bar{u})$ denote the effective 
M2-brane 
charge depending on $\bar{u}$, and scale factor, respectively,
\Eq{
H(\bar{u}, r)=1+\frac{\bar{M}(\bar{u})}{r^6}\,,~~~~~
a_{\rm M2}(\bar{u})=\left({\bar{u}\over \bar{u}_0}\right)^{3/2}
\,,
\label{p}
}
with
\Eqrsubl{BH11:M:Eq}{
\bar{M}\left(\bar{u}\right) 
&\equiv& \left({\bar{u}\over \bar{u}_0}\right)^{-3}M\,,\\
cu&=&\left(\frac{\bar{u}}{\bar{u}_0}\right)^3\,,~~~~~
\bar{u}_0 \equiv  {3\over c}\,.
}

The near M2-brane geometry is the same as the static one because 
there is a null Killing vector at the horizon 
and then the geometry approaches the static solution. 
Since it has a horizon geometry, we can regard
the present dynamical solution as a black hole.
The dynamical M2-brane gives the black hole spacetime  
while the asymptotic structure in the dynamical M2-brane 
is completely different from that of a static one.
Although the static M2-brane solution has an asymptotically flat geometry, 
the dynamical M2-brane solution is a time dependent 
anisotropic spacetime at a constant $x$ coordinate.

%%%%%%%%%%%%%%%%%%%%%%%%%%%%%%%%%%%%%%%%%%%%%%%%%%%%%%%%%%
%%%%%%%%%%%%%%%%%%%%%%%%%%%%%%%%%%%%%%%%%%%%%%%%%%%%%%%%%%
%T2>Black hole in 10-dimensional effective theory
\subsection{Black hole in the ten-dimensional effective theory}
%%%%%%%%%%%%%%%%%%%%%%%%%%%%%%%%%%%%%%%%%%%%%%%%%%%%%%%%%%
%%%%%%%%%%%%%%%%%%%%%%%%%%%%%%%%%%%%%%%%%%%%%%%%%%%%%%%%%%
In this section, we study the dynamics of the M2-brane black hole in the 
lower-dimensional background after compactifying the internal space. 
Now we compactify a one-dimensional 
M2-brane world volume just as the case of a static black 
hole and consider the ten-dimensional effective theory. 
In this case, we find the 11-dimensional metric
\Eq{
ds^2=ds_{10}^2+ds_1^2\,,
} 
where
\Eqrsubl{ef:metric:Eq}{
ds_{10}^2&=&h^{-2/3}(u, r)\left[-2dudv+h(u, r)\left(dr^2
+r^2d\Omega_{(7)}\right)\right],\\
ds_{1}^2&=&h^{-2/3}(u, r)(dy)^2\,. 
}
The compactification of $ds_{1}^2$ gives the effective ten-dimensional
spacetime, whose metric in the Einstein frame $d\bar s_{10}^2$ is given by
\Eq{
d\bar s_{10}^2=h^{-3/4}(u, r)\left[-2dudv+h(u, r)\left(dr^2
+r^2d\Omega_{(7)}\right)\right].
  \label{ef:Emetric:Eq}
}
If we use a new coordinate $\tilde{u}$
\Eq{
cu=\left(\frac{\tilde{u}}{\tilde{u}_0}\right)^4\,,~~~~\tilde{u}_0=\frac{4}{c}
\,,
}
the ten-dimensional metric (\ref{ef:Emetric:Eq}) can be rewritten explicitly as
\Eqr{
d\bar s_{10}^2=\tilde{H}^{-3/4}(\tilde{u}, r)
 \left[-2d\tilde{u}dv+a_{\rm eff}^2(\tilde{u})\,\tilde{H}(\tilde{u}, r)\,
\left(dr^2+r^2d\Omega_{(7)}\right)\right],
\label{ef:Emetric2:Eq}
}
where the function $\tilde{H}(\tilde{u}, r)$ is given by 
\Eq{
\tilde{H}(\tilde{u}, r)=1+{\tilde{M}(\tilde{u})\over r^6}\,,
}
with the effective M2-brane charge $\tilde{M}(\tilde{u})$\,, 
and the scale factor 
$a_{\rm eff}(\tilde{u})$\,,
\Eqrsubl{ef:par:Eq}{
\tilde{M}(\tilde{u}) &\equiv& \left({\tilde{u}\over \tilde{u}_0}\right)^{-{4}}M\,,\\
a_{\rm eff}(\tilde{u})&=& \left({\tilde{u}\over \tilde{u}_0}\right)^2\,.
}

From Eqs.~(\ref{ef:Emetric:Eq}), (\ref{ef:Emetric2:Eq}) 
in the limit of $r\rightarrow \infty$,
we find
\Eqr{
d\bar s_{10}^2&=&(cu)^{-3/4}\left[-2dudv+cu\left(dr^2
+r^2d\Omega_{(7)}\right)\right]
\nonumber\\
&=&-2d\tilde{u}dv+a_{\rm eff}^2(\tilde{u})\, \left(dr^2
+r^2d\Omega_{(7)}\right)\,.
}
Since our solution approaches an asymptotically time dependent universe
with the scale factor $a_{\rm eff}(\tilde{u})$\,, we can regard again 
the time-dependent M2-brane solution as a black hole in the expanding 
Universe. 
If we compactify the direction of the world volume coordinate,
we find the different power exponent of time in the scale factor,
which is also shown in the original 11-dimensional background.

As a result, we always find the different power of time $\tilde{u}$ 
in the scale factor $a_{\rm eff}(\tilde{u})$ 
for a $d$-dimensional black hole $(d\le 10)$ if we smear the 
transverse space to the M2-brane. If $d_{\rm s}$-dimensions of the transverse 
space to the M2-brane are smeared, 
which gives the different power of 
transverse space coordinates to the M2-brane 
\Eqrsubl{ef:sm:Eq}{
ds^2&=&h^{-2/3}(\tilde{u}\,, z)\left[-2d\tilde{u}dv+\left(dy\right)^2
+h(\tilde{u}\,, z)\,\delta_{ab}dz^adz^b\right],\\
h(\tilde{u}\,, z)&=& c \tilde{u}+\sum_l
\frac{M_l}{|z^a-z^a_l|^{6-d_{\rm s}}}\,,~~~~~(d_{\rm s}\le 7)
}
in terms of the multi-black-hole coordinates. 
Here, $z^a~(a=1, 2, \cdots , 8)$ 
denote the coordinates of the transverse space to the M2-branes, 
$M_l~(l=1, 2, \cdots , N)$ are M2-brane charges, 
and $z^a_l~(l=1, 2, \cdots , N)$ are positions of M2-branes. 
Suppose $d_{\rm s}$ dimensions of the transverse space to M2-branes
 are smeared 
and compactified, where $d_{\rm s}\le 7$\,. If one compactifies 
the $d_{\rm s}$-dimensional transverse space as well as 
the $d_{\rm M}(=0~{\rm or}~1)$-dimensional 
M2-brane world volume, the 
$d\left[=(11-d_{\rm M}-d_{\rm s})\right]$-dimensional metric in 
the Einstein frame 
is given by 
\Eqrsubl{ef:bg:Eq}{
&&\hspace{-1cm}d\bar s_{d}^2=H^{\frac{d_{\rm s}-6}{d-2}}
(\tilde{u}, z)\left[-2d\tilde{u}dv+\left(1-d_{\rm M}\right)
a_{\rm eff}^{\frac{2(d_{\rm s}-6)}{d-2}}
\left(dy\right)^2+a_{\rm eff}^2(\tilde{u})\,H(\tilde{u}, z)
\,\delta_{PQ}dz^Pdz^Q\right],\\ 
&&\hspace{-1cm}
H(\tilde{u}, r)=1+{\tilde{M}(\tilde{u})\over r^6}\,,
}
where $\delta_{PQ}dz^Pdz^Q$ is the metric of 
$\left(8-d_{\rm s}\right)$-dimensional Euclidean space. 
The effective M2-brane charge $\tilde{M}(\tilde{u})$, 
the scale factor $a_{\rm eff}(\tilde{u})$\,, and a coordinate $\tilde{u}$
are also given by
\Eqrsubl{ef:smear:Eq}{
&&\tilde{M}(\tilde{u}) = \left({\tilde{u}\over \tilde{u}_0}\right)^
{-\frac{d-2}{3-d_{\rm M}}}M\,,~~~~
a_{\rm eff}(\tilde{u})= \left({\tilde{u}\over \tilde{u}_0}\right)^
{\frac{d-2}{2(3-d_{\rm M})}}\,,\\
&&cu=\left(\frac{\tilde{u}}{\tilde{u}_0}\right)^
{\frac{d-2}{3-d_{\rm M}}}\,,
~~~~\tilde{u}_0=\frac{d-2}{c(3-d_{\rm M})}\,.
}
This power exponent is obtained for an universe filled by the 
four-form field strength satisfying the field equation. 
We may regard the present 
$d$-dimensional solution as a time-dependent black hole.

%======================================%
%<<<<<<<<<<<<< SECTION 6 >>>>>>>>>>>>>>%
%======================================%
%T1>Discussions
\section{Discussions}
  \label{sec:discussions} 
In the present paper, we have constructed the dynamical 
supersymmetric M2-brane solution for the warped compactification of 
an 11-dimensional supergravity. 
The solution is given by an 
extension of a static supersymmetric M2-branes solution.  
In the case of a dynamical M2-brane background, 
a quarter of maximal supersymmetries exists. 
If the M2-brane charge vanishes, our solution gives a plane wave background 
which preserves a half of the full supersymmetry. Therefore, 
in the far region from the M2-brane, 
the background changes from the dynamical M2-brane to 
time-dependent plane wave background. 
This means that one quarter of the maximal supersymmetry is 
enhanced to a half of the possible rigid supersymmetries  
in the maximal case when one moves in the 
transverse space to the M2-brane. 
Although we have mainly discussed the single M2-brane 
solution in this paper, it is possible to generalize it to the 
solution which describes an arbitrary number of extremal M2-branes in 
an expanding universe. 
We have found that the degree of the supersymmetry breaking is 
strongly related to the dynamics of the background. Then, the time evolution 
of the geometry is deeply connected with the hierarchy and 
supersymmetry breaking while the inhomogeneity of the M2-brane 
world volume coordinates makes preserving the supersymmetry.  
In the region where the effect of the inhomogeneity of 
the M2-brane world volume coordinates is  
smaller compared to 
the contribution of the M2-brane charge, our supersymmetric solution 
describes the breaking of the supersymmetry, which is the transition from 
the supersymmetric universe to a nonsupersymmetric one as time evolves. 

The dynamical M2-brane solutions can always take a form in the function 
$h(x, r)=h_0(x)+h_1(r)$\,, where the function  
$h(x, r)$ depends on the linear function of the 
M2-brane world volume coordinates $x^{\mu}$ as well as coordinates of the 
transverse space to the M2-brane\,. Since the existence of 
the function $h_0(x)$ implies the dynamical instability in the moduli of 
internal space \cite{Kodama:2005fz}, it would be useful to study the stability 
of a solution. 

Motivated by the construction of a new supersymmetric solution, we have 
studied the global structure of the dynamical M2-brane background. 
We have found that the time dependence changes the causal structure 
of a static M2-brane solution. 
Since the volume element of any congruence along the radial 
null geodesic vanishes at the curvature singularity, 
it turns out that this is a strong version of a timelike singularity. 
We have studied null geodesics which terminate a coordinate singularity
in terms of an analytic extension across there and showed that there is a 
regular null hypersurface (or horizon) generated by a 
null Killing vector field.
In particular, this null Killing vector field describes the generator of 
the horizon even if the bulk metric is asymptotically 
anisotropic geometry 
at a constant $x$ coordinate. 
Hence, the near horizon geometry in this solution 
gives the regular spacetime, and thus becomes AdS${}_4\times$S${}^7$\,. 

It is important to explore another analytic solution describing a 
supersymmetric 
M-brane or D-brane in the expanding Universe. One may present whether 
supersymmetric dynamical brane solutions affect the formation of the naked 
singularity. 
Upon setting an appropriate initial condition, these solutions may allow 
us to violate the cosmic censorship \cite{Horne:1993sy, Maeda:2015joa}.

We can also discuss a dynamical black hole solution whose spacetime gives 
a time dependent universe. The near M2-brane region of this black hole 
in the expanding Universe is the same as the static solutions while 
the asymptotic structures are completely different, giving the 
anisotropic spacetime at a fixed $x$ coordinate with scale factors 
for a dynamical universe. The effective M2-brane charge for the 
supersymmetric background depends on the world volume coordinates of 
the M2-brane. 

The supersymmetric solutions can contain the function depending on null  
coordinates of the M2-brane world volume direction. 
The results we have obtained are not unnatural because studies of 
the supersymmetric plane wave background 
 showed that it is possible to obtain time-dependent supersymmetric solutions 
with a nontrivial dependence on spacetime coordinates 
\cite{Blau:2001ne, Sakaguchi:2003cy}. 
Although this may be a limitation on the applications of our solution, 
it is interesting to explore if similar more general dynamical and 
supersymmetric solutions can be obtained by relaxing or extending 
some of our assumptions for the 10-, 11-, or lower-dimensional backgrounds. 
We will study this subject in the near future.

%T1>Acknowledgments
\section*{Acknowledgments}
We thank Akihiro Ishibashi for careful reading of the manuscript and valuable 
comments. K. U. would like to thank Masato Nozawa for useful comments.  
K. U. also thank the Yukawa Institute for Theoretical Physics at Kyoto University for hospitality during the YITP workshop on "Gravity and Cosmology for Young Researchers" (Workshop No. YITP-X-16-10) which was supported by Grant-in-Aid for Scientific Research on Innovative Areas No. 15H05888 "Multifaceted Study of the Physics of the Inflationary Universe". This work is supported by Grants-in-Aid from the Scientific Research Fund of the Japan Society for the Promotion of Science, under Contracts No. 17K05451 (K. M.) and No. 16K05364 (K. U.).

%======================================%
%<<<<<<<<<<<<< REFERENCE >>>>>>>>>>>>>>%
%======================================%

%T1>References

\end{document}